\documentstyle[11pt]{article}
 
\begin{document}

\baselineskip=14pt plus 0.2pt minus 0.2pt
\lineskip=14pt plus 0.2pt minus 0.2pt

\newcommand{\be}{\begin{equation}}
\newcommand{\ee}{\end{equation}}
\newcommand{\bea}{\begin{eqnarray}}
\newcommand{\eea}{\end{eqnarray}}
\newcommand{\da}{\dagger}
\newcommand{\dg}[1]{\mbox{${#1}^{\dagger}$}}
\newcommand{\hlf}{\mbox{$1\over2$}}
\newcommand{\lfrac}[2]{\mbox{${#1}\over{#2}$}}
\newcommand{\scsz}[1]{\mbox{\scriptsize ${#1}$}}
\newcommand{\tsz}[1]{\mbox{\tiny ${#1}$}}

\begin{flushright}
quant-ph/9908059 \\
LA-UR-99-4097 \\
\end{flushright} 

\begin{center}
\Large{\bf 
Electrons above a Helium Surface \\ 
and the One-Dimensional Rydberg Atom} \\
\vspace{0.25in}

\large
\bigskip

Michael Martin Nieto\footnote{\noindent  Email:  
mmn@lanl.gov}\\
{\it Theoretical Division (MS-B285), Los Alamos National Laboratory\\
University of California\\
Los Alamos, New Mexico 87545, U.S.A. \\}

\normalsize

\vskip 20pt
\today

\vspace{0.3in}

{ABSTRACT}
 
\end{center}
\begin{quotation}
\baselineskip=.33in

Isolated electrons resting above a helium surface are predicted 
to have a bound spectrum corresponding to 
a one-dimensional hydrogen atom.  But in fact, the observed spectrum is 
closer to that of a quantum-defect atom.  Such a model is discussed and 
solved in analytic closed form. 

\vspace{0.25in}

\noindent PACS: 03.65.Ge, 73.20-r

\end{quotation}

\newpage

\baselineskip=.33in

Some time ago the prediction was made 
that an isolated electron  resting on a helium (or some certain other) 
surface should have 
a bound-state spectrum in the vertical direction \cite{ta}-\cite{tc}. 
The idea is that the electron induces an image 
charge in the helium, producing a potential on the electron of 
\bea
V(x) &=& - \frac{Z e^2}{x}  , 
                      ~~~~~~~x>0, 
             ~~~~~~~~~~Z = \frac{(\epsilon -1)}{4(\epsilon+1)},
\\
     &=& + \infty,~~~~~~~~~~x\le0,   
\eea  
where $\epsilon$ is the dielectric constant \cite{LL}.  For  helium it is 
\cite{epsilon, a}
\be
\epsilon = 1.05723,~~~~~~~~~Z=0.0069547.
\ee
The spectrum should thus be similar 
to that of a (weakly-coupled) one-dimensional hydrogen atom.  
This phenomena has been observed \cite{a}-\cite{b}. 
(See \cite{d} for a current review and \cite{qcomp} for a proposed 
application to quantum computing.)

Consider the 
one-dimensional Schr\"odinger equation of this system  \cite{1DH}:
\be
\left(-\frac{\hbar^2}{2m}\frac{d^2}{dx^2} 
         - \frac{Ze^2}{x}\right)\psi_n(x) = E_n\psi_n(x).
\ee
Making the changes of variables 
\bea
E_n = -\frac{{\cal E}_0}{n^2},& ~~~~~& 
               {\cal E}_0 = \frac{mZ^2e^4}{2\hbar^2},  \\
z_n = \frac{x}{n x_0}, &~~~~~& x_0=\frac{\hbar^2}{2mZe^2},
\eea
one obtains 
\be
\left(\frac{d^2}{dz_n^2} + \frac{n}{z_n} -\frac{1}{4}
\right)\psi_{n} = 0.
\ee
Observe that the helium-surface ``Rydberg'' and ``Bohr radius'' have values
\bea
{\cal E}_0 &=&  Z^2R_{\infty} = 0.658086~meV=159.123~GHz,   \\
      b_0    &=& 2 x_0 = a_0/Z = 76.01~\AA   .
\eea 

By techniques similar to those used to obtain 
the solutions for the  3-dimensional hydrogen atom, 
one can obtain the normalized eigensolutions \cite{mmnND,J}
\bea
\psi_{n}(z)& =&  [2n^3 x_0]^{-1/2} ~z_n~\exp[-z_n/2] ~
                        L_{n-1}^{(1)}(z_n).    \label{psiN}
\eea
This agrees with particular $n=1,2,3$ wave functions in the 
literature \cite{psione}.

In Eq. (\ref{psiN}) 
we have used the {\it generalized} Laguerre polynomials commonly 
found in the modern mathematical physics literature \cite{mos}: 
\be
L_n^{(\alpha)}(x)= \sum_{k=0}^n
\left(   \begin{array}{c}
          n+\alpha \\
           n-k        \end{array}   \right)
  \frac{(-x)^k}{k!}
                =\frac{e^x x^{-\alpha}}{n!}~\frac{d^n}{dx^n}
                   \left[e^{-x}x^{n+\alpha}\right]. \label{Lus}
\ee
In Eq. (\ref{Lus}), the generalized binomial symbol 
     $ (   \begin{array}{c} a \\ b  \end{array} )$ 
means $\Gamma(a+1)/[\Gamma(a-b +1)\Gamma(b+1)]$.
Also, $L_n^{(0)}(x)= L_n(x)$,  where $L_n(x)$ are the ordinary Laguerre 
polynomials normalized to unity at zero: $L_n(0)=1$.  
These polynomials 
were used instead of the {\it associated} Laguerre polynomials often 
defined, for Coulomb wave functions \cite{schiff}, as
\be
L_n^j(x) \equiv \frac{d^j {\bar L}_n(x)}{dx^j} 
=\frac{d^j}{dx^j}\left[
    e^x \frac{d^n}{dx^n}\left(e^{-x}x^n\right)\right] . 
\label{Lord}
\ee
Here, ${\bar L}_n(x)=(n!)L_n(x)$, the ordinary Laguerre polynomials 
normalized to ${\bar L}_n(0)=(n!)$.   
Eq. (\ref{Lord}) can be confusing, since this definition only holds 
for integer $j$.  Contrariwise,  Eq. (\ref{Lus}) is defined for 
arbitrary $\alpha$.  (This will be very important in the following.) 
When $\alpha = j$, an integer, the connection between the two forms is 
\be
L_{n+j}^j(x) = (-1)^j [(n+j)!]~L_n^{(j)}(x).
\ee  

The experiments obtain transition energies from excited states to 
the ground state:   
\be
\Delta_n = |E_{1}| - |E_n| ={\cal E}_0\left(1 - \frac{1}{n^2}\right).
\ee
However, the experiments do not yield exact Balmer energies.  
The $\Delta_n$ are all of order 7 GHz too large \cite{a}-\cite{c}.  
This is like a quantum defect, since if 
\be
E_n \rightarrow E_{n^*}= - \frac{{\cal E}_0}{(n^*)^2},
~~~~~~~~n^*=n-\delta,
\label{En*} 
\ee
then 
\bea
\Delta_{n^*}& = & |E_{1^*}| - |E_{n^*}| 
      ={\cal E}_0\left(\frac{1}{(1-\delta)^2}-\frac{1}{(n-\delta)^2}\right)\\
&\approx& {\cal E}_0\left[\left(1-\frac{1}{n^2}\right)  
        + 2\delta\left(1-\frac{1}{n^3}\right) 
        + 3\delta^2\left(1-\frac{1}{n^4}\right) + \dots\right].
\eea
The correction term, ${\cal E}_0~2~\delta~[1-{1}/{n^3}]$, 
varies by only $\sim 10$ \% as $n$ varies from $2$ to $\infty$ \cite{ref1}.
Just fitting the $2^* \rightarrow 1^*$ and $3^* \rightarrow 1^*$
transition energies \cite{b} to this 
formula yields, ${\cal E}_0 = 158.4$ GHz and $\delta = 0.0237$ or 
an increase in $\Delta$ of about 7.8 GHz.  
In other words, this is like a one-dimensional Rydberg atom.  

Elsewhere \cite{r1}-\cite{r4}, 
inspired by supersymmetry \cite{r0}, it was shown how one can obtain  exact, 
analytic, one-particle wave functions for real Rydberg atoms   
yielding the correct eigenenergies.  
This also yielded: (a) transition matrix elements in agreement with 
experiment and complicated many-body calculations \cite{r1}; 
(b) good fine-structure splittings \cite{r2};  and (c) Stark splittings 
whose crossing/anti-crossing patterns agree with experiment \cite{r3}.  

The mathematical key to this success is the fact that 
for proper solutions of the (radial) hydrogen-atom equation 
one does not really need that $l=$(integer) and $n=$(integer).  
One only needs that $(n-l)=$(integer), the two 
separately not having to be integers.  That is, the factor $l(l+1)$ 
in the effective $1/r^2$ potential term need not have $l$ be an 
integer for a finite-order polynomial radial solution to exist.  
This is where  the $L_n^{(\alpha)}(x)$ become of use \cite{qdold}.

Applying this idea to the present case, we phenomenologically propose 
for $x>0$ that $V(x)$ becomes 
\be
V(x) = - \frac{Ze^2}{x} 
       +\frac{\hbar^2}{2m}\frac{(-\delta)[(-\delta)+1]}{x^2},
~~~~~~~x>0.
\ee
Then the exact eigenenergies are given by Eq. (\ref{En*}) and the 
exact wave functions are 
\bea
\psi_{n^*}(z) & =& N_{n^*}
                        ~z_{n^*}^{1-\delta}~\exp[-z_{n^*}/2] ~
                        L_{n-1}^{(1-2\delta)}(z_{n^*}), \\
N_{n^*}&=&\left[\frac {1}{ 2(n^*)^2x_0}~ 
            \frac{\Gamma(n)}{\Gamma(n+1-2\delta)}\right]^{1/2}, 
         ~~~~~~~~~~~~~z_{n^*} = \frac{x}{n^*x_0}.
\eea
Thus, we have an exact analytic solution to the problem.  We can also 
analytically calculate the 
expectation values $\langle j^*| x^t|k^* \rangle$ \cite{ref2} as  
double sums of Gamma functions \cite{buch}.  In particular \cite{J},
\be
\langle x \rangle_{n^*} 
        =  x_0\left[3n^2 -\delta(6n-1-2\delta)\right]. \label{nxn}
\ee
When $\delta=0$ this reduces to the standard result.  Also , 
\bea
\langle 1^*| x|n^* \rangle 
   &=& \frac{x_0 g^{4-2\delta}}{2}\left(\frac{n^*}{1^*}\right)^\delta
    \left[\frac{\Gamma(n+1-2\delta)\Gamma(n)}
                 {\Gamma(2-2\delta)}\right]^{1/2}  \nonumber \\
  &~& \times ~ \sum_{k=0}^{n-1}\frac{(-g)^k}{k!}
              ~\frac{(k+3-2\delta)(k+2-2\delta)}{(n-1-k)},
     ~~~~~~~g =\frac{2\cdot 1^*}{n^* + 1^*}.
\label{1xn}
\eea 
Setting $n^*$ to $1^*$ in Eqs. (\ref{nxn}) and  (\ref{1xn}), makes 
then equal.     

Unfortunately, this model  
does not resolve the physical problem of how one realistically cuts off 
the unphysical, negatively-infinite potential at the origin \cite{cutoff}. 
In fact, this solution makes the problem slightly worse:  at the origin 
the potential now goes to negative infinity 
as $-1/x^2$.  If the experimental quantum defect had been 
of opposite sign, then the added potential would have been positive, 
like an angular momentum barrier, making the states less bound.  
This also would have `` realistically'' modeled 
the positive work function at the surface of about 1 eV \cite{b}.  

There are inverse methods for generating inequivalent isospectral 
Hamiltonians \cite{AM}-\cite{P}.  What, in principle, would be 
an isospectral Hamiltonian with the desired physical properties 
is one with an added potential  that: 
(i) goes, at the origin, to plus infinity at least  slightly faster  
than $  [\delta(1-\delta)]/z^2$,    
(ii) becomes negative for larger $z$, and  
(iii) goes to zero at infinity from below.  

A first examination of the above inverse methods 
\cite{AM}-\cite{P}  found 
potentials with the last two properties, but not the first. These 
potentials go  to zero at the origin.  An example is \cite{AM,P}
\bea
V_2(z)& = & \frac{2}{(1^*)^2}
\left[
\left(\frac{2-2\delta}{z_{1^*}} -1\right) 
Y + Y^2 
\right],   ~~~~~ x>0,
\\
Y & = &  \left[\frac{\exp[-z_{1^*}]~z_{1^*}^{2-2\delta}}
        {\Gamma(3-2\delta,z_{1^*}) - R}\right],    
\eea
where $R$ (which can be chosen to be $-2$) 
and $\Gamma(a,z)$ (the incomplete $\Gamma$ function) are 
\be
R \equiv \frac{\gamma +1}{\gamma~ (1^*)~ N_{1^*}^2}
   =  \frac{\gamma +1}{\gamma}\Gamma(3-2\delta), ~~~~~~~~~~~
\Gamma(a,z) = \int_z^{\infty}dy~ y^{a-1}~e^{-y}, 
\ee
and $\gamma$ is a dependent constant useful below.  Taking units of 
$x_0=1$, the orthonormal eigenfunctions are 
\bea
\chi_{n^*}(z)&=& \psi_{n^*}(z) +\int_0^z dy~K(z,y)~\psi_{n^*}(y),
             ~~~~~~~  n >1, \\
K(z,y)&=&\left( \frac {1}{1^*} \right)
   \frac{\exp[-z_{1^*}/2]~\left(z_{1^*}^{1-\delta}\right)~
         \exp[-y_{1^*}/2]~\left(y_{1^*}^{1-\delta}\right) }
{\Gamma(3-2\delta,z_{1^*}) - R}. 
\eea
The  exception, 
with normalization $\gamma/(\gamma + 1)$, is
\be 
\chi_{1^*}(z) = \left( \frac {-\Gamma(3-2\delta)}{\gamma^{1/2}} \right)
\frac{\psi_{1^*}(z)}
          {\Gamma(3-2\delta,z_{1^*}) - R}.
\ee
There may well be analytic isospectral Hamiltonians with all the desired 
properties. But to determine their existence requires more investigation.  


\noindent {\bf Acknowledgements}

I wish to thank C. C. Grimes,  D. K. Lambert, 
T. Opatrny, P. L. Richards, L. Spruch, 
and especially P. M. Platzman for very helpful comments on the 
physics of electrons on a helium surface. V. A. Kosteleck\'y 
and P. W. Milonni kindly reviewed a draft.  
The support of the United States Department of 
Energy is acknowledged.



\begin{thebibliography}{99}
\bibitem{J} To obtain this result, the integral (shown in Ref. 
\cite{mmnND}) 
\bea
 J^{(\beta)}_{n,\alpha} &=& 
\int_0^\infty dt \exp[-t] t^{\alpha+\beta} [L^{(\alpha)}_n(t)]^2 
\nonumber \\
&=& \frac{ \Gamma(n+1+ \alpha)}{\Gamma(n+1)]}
      \sum_{j=0}^n 
\frac{(-1)^{j-n} \Gamma(j+1+ \alpha+ \beta)]}{\Gamma(j+1) \Gamma(j+1+ \alpha) 
       \Gamma(n+1-j)}
\frac{ \Gamma(j+1+\beta)}{ \Gamma(j+1+ \beta-n)]}. 
\eea
is useful.  Note, in particular, that when 
$\beta= (n - j -1)$ is  an integer, t
then the last $\Gamma$ function  in the denominator cuts off the sum.
\end{thebibliography}
\end{document}